\documentclass[aps,pre,twocolumn,groupedaddress]{revtex4}
\usepackage{graphicx}
\def\be{\begin{equation}}
\def\ee{\end{equation}}
\begin{document}
\draft

\title{Standard map in magnetized relativistic systems: fixed points and regular acceleration}
\author{M.C. de Sousa\footnote{meirielenso@yahoo.com.br}, F.M. Steffens,}
\address{Universidade Prebisteriana Mackenzie, Departamento de F\'{\i}sica - NFC 
Rua da Consola\c{c}\~ao 930, 01302-906 S\~ao Paulo, SP, Brasil}
\author{R. Pakter, F.B. Rizzato}
\address{Instituto de F\'{\i}sica,
Universidade Federal do Rio Grande do Sul\\
Caixa Postal 15051, 91501-970, Porto Alegre, RS, Brasil}

\begin{abstract}
We investigate the concept of a standard map for the interaction of relativistic particles and electrostatic waves of arbitrary 
amplitudes, under the action of external magnetic fields. The map is adequate for physical settings where waves and 
particles interact impulsively, and allows for a series of analytical result to be exactly obtained. Unlike the traditional form of the 
standard map, the present map is nonlinear 
in the wave amplitude and displays a series of peculiar properties. Among these properties we discuss the relation involving 
fixed points of the maps and accelerator regimes.
\end{abstract}

\maketitle

\section{Introduction}

Wave-particle interaction has always attracted a great deal of attention as an 
efficient way to particle acceleration and particle heating. A wide range 
of applications indeed flourishes, from heating and current drives in fusion 
devices \cite{Karney78, Fisch87}, to more alternative concepts like plasma based particle 
accelerators \cite{Tajima79} and non-neutral beams \cite{Davidson01}. 

Wave-particle interaction is basically a nonlinear process \cite{Shukla86,Mendonca2001}. 
This means that in the usual scenarios of the interaction one can expect to see 
regular and chaotic patterns interwined in the appropriate phase-spaces \cite{Lichtenberg92}. Regular 
regions are useful for coherent acceleration while chaotic regions are adequate for 
particle heating. The prevalence of one or another type of region is a direct 
result of the strength of the perturbations impinging on the the particle motion. 

As shown over the years, investigation of nonlinear wave-particle interaction can be largely aided if area 
preserving Hamiltonian maps can be constructed for the system under study. Area preserving maps provide such a powerful 
tool in the study of the nonlinear dynamics of Hamiltonian systems because they yield a series of exact 
analytical results to work with \cite{Lichtenberg92, Caldas2009}.

Among the large variety of area preserving maps, the most prominent is perhaps the standard canonical map in 
one-degree-of-freedom and its close variants \cite{Corso98,Prado2000}. The standard map describes a large number 
of systems, among which one can find the case of wave-particle nonlinear interaction where particles move 
under the action of electrostatic fields \cite{Shukla86,Mendonca2001}. Under these circumstances the map clearly 
shows how appropriate resonant conditions give rise to efficient acceleration mechanisms.

Maps can be constructed from the underlying Hamiltonians as one makes the assumption that 
the system is periodically perturbed by impulsive kicks, followed by integration of the resulting canonical equations 
over one perturbing period. Impulsive approximations adequately represent pulsed perturbations, resulting from 
broad band spectra of counter propagating waves driven by nonlinear wave coupling \cite{Chernikov89,Nomura92}.

In the standard case, while one of the canonical variables appears only in the interaction 
term of the original Hamiltonian, the other appears only in the free part. This peculiar feature allows to show that 
one of the variables is constant between the kicks and the other does not vary across a kick, which is the key factor 
to simplify the final form of the theory. As a further result, the map can be shown to be linear in the wave amplitude. 
If one now takes another case of interest where particles move under the action of an external magnetic field, 
the canonical variables become entangled by the field and can no longer be easily segregated into the 
free and interaction terms of the corresponding Hamiltonian. 
However, as will be shown, a set of convenient variable changes allow for the construction of an 
exact map. 

We shall focus on electrostatic perturbing modes propagating perpendicularly to the external 
magnetic field. These modes are frequently present and can be responsible for a large amount of particle energization. 
In open boundary systems, traveling electrostatic waves have been shown to produce efficient coherent 
acceleration \cite{Karney78, Corso93,Choueiri04} as well as incoherent heating \cite{Yoon06}. 
In closed systems, stationary modes formed with counter propagating waves 
are the ones to be considered. Closed systems encompass the cases of magnetically confined plasma columns and 
beams, and support waves with frequencies in the vicinity of the cyclotron frequency \cite{Krall73,Lund98}. Resonant 
cyclotronic interaction is also possible between lower hybrid modes and fast electrons when the electronic cyclotron frequency 
decreases due to relativistic effects \cite{Rax91}. 
Large resonant islands \cite{Pakter94} will be shown to form due to relativistic nonlinearities of the transverse 
particle motion with respect to the magnetic field, a feature of relevance in accelerator regimes as discussed later on. 

The central interest of the present paper is then to 
construct the exact map for the dynamics of magnetized particles under the action of perpendicular electrostatic waves. 
The resulting map could be seen as the magnetized counterpart of the classical standard map. Although 
related, we shall see that both maps display very different structures: nonlinearities in the dependence on the 
wave amplitude affects positioning of fixed points and the associated accelerating regimes of the 
magnetized case. Limits on the acceleration efficiency of the magnetized case are discussed as well. 

The paper is organized as follows: \S 2 is devoted to the construction of the map, \S 3 to the analytical and 
numerical investigation, and in \S 4 we draw our conclusions.

\section{From the model to the nonlinear map}

Consider a particle with charge 
$q$, mass $m$ and perpendicular canonical momentum ${\bf p}_{\perp}$, moving under the 
combined action of a uniform magnetic field ${\bf B} = B_0 \hat {\bf z}$ and a stationary 
electrostatic wave of wavevector $k$, period $T$ and amplitude $A$, lying along the $x$ axis. Working with a set of 
dimensionless quantities, 
${\bf p}_{\perp}/mc \rightarrow {\bf p}_{\perp}$, $q B_0 x/mc \rightarrow x$, $(q B_0/m)(t,T) \rightarrow (t,T)$, 
$(1/mc^2) (q B_0/m)\,A\rightarrow A$,  the dimensionless Hamiltonian $H \rightarrow H/mc^2$ 
governing the particle motion can be written as
\begin{equation}
H = \sqrt{1+p_x^2+x^2} + A \cos (k x) \sum_n \delta \left(t - n \, T\right),
\label{equa3}
\end{equation}
where with no loss of generality we consider the canonical momentum $p_y = 0$. We point out that even though the 
y-component of the canonical momentum is conserved and taken to be zero, $dy/dt$ is not zero and the motion is 
not one-dimensional. $c$ is the speed of light and, as stated earlier, for our purposes we focus on pulsed systems 
whose action is represented by the periodic collection of delta functions. 

At this point one must adopt a strategy to integrate the dynamics generated by the Hamiltonian 
(\ref{equa3}). To this end, we note that particles are subjected to kicks whenever the time is a multiple of the period $T$.
Between consecutive kicks, however, the one-degree-of-freedom dynamics is time independent and 
integrable, thus representable in terms of action-angle variables $p_x = \sqrt{2 I} \cos \theta,\>\>
x= \sqrt{2 I} \sin \theta$ with constant action $I$. For future purposes, one can promptly find the form of the 
Hamiltonian written in terms of action-angle coordinates. It reads
\begin{equation}
H=\sqrt{1+ 2 I} + A \cos(k \sqrt{2 I} \sin \theta) \sum_n \delta \left(t - n \, T\right).
\label{equa3aa}
\end{equation}

Across the kick, both $I$ and $\theta$ undergo abrupt variations since both 
are present in the wave-particle coupling term. However, in the original variables $x,\>\>p_x$ only $p_x$ 
changes across the kick - $x$ remains constant because $p_x$ is absent from the coupling. That said, our approach is the following:

\begin{enumerate}

\item[(i)] we first define the action-angle variables just before the $n^{th}$ kick: $I_n,\theta_n$;

\item[(ii)] action and angle are then transformed to 
the original cartesian variables via 
$(x_n,\>\>{p_x}_n) = (\sqrt{2I_n} \sin \theta_n, \>\> \sqrt{2 I_n} \cos \theta_n)$; 

\item[(iii)] changes across kick $n$ are calculated in the cartesian coordinates via  
$\Delta x = 0, \Delta p_x = k A \sin (k x_n)$; 

\item[(iv)] immediately after the kick we write ${p_x}_n^+ \equiv {p_x}_n + \Delta p_x$ and $x_n^+ = x_n + \Delta x$ and 
change back to action-angle variables via 
$I_n^+ = ({{p_x}_n^+}^2 + {x_n^+}^2)/2$, $\theta_n^+ = \arctan \left(x_n^+ / {p_x}_n^+ \right)$. 

\item[(v)] The last step is to propagate 
the preceding state to that instant $n+1$ before the next kick via 
$I_{n+1} = I_{n+1}^+,\>\> \theta_{n+1} = \theta_n^+ + T/\sqrt{1+2 I_{n+1}}$. 

\end{enumerate}
Although the final composite map could be expressed with similar degree of complexity 
in terms of cartesian coordinates, use of action-angle variables is more convenient in view of the 
fact that action is conserved in the absence of perturbation.

The final result can be cast in the form 
of an explicit map relating the dynamical states at kicks $n$ and $n+1$:

%
%
\begin{widetext}
\begin{eqnarray}
I_{n+1} = {1 \over 2}  \left(2 I_n \sin^2 \theta_n + \left(\sqrt{2 I_n} \cos \theta_n + {1 \over 2} \varepsilon 
k \sin(k \sqrt{2 I_n} \sin \theta_n)\right)^2\right), \label{equa4a}\\ \cr \cr \cr
\theta_{n+1} = \arctan \left( {2 \sqrt{2 I_n} \sin \theta_n \over 2 \sqrt{2 I_n} \cos \theta_n + 
\varepsilon k \sin(k \sqrt{2 I_n} \sin \theta_n)} \right) + {T \over \sqrt{1+2 \, I_{n+1}}},
\label{equa4b}
\end{eqnarray}
\end{widetext}
where for convenience we introduce $\varepsilon \equiv 2\,A$. The map has the noticeable feature of being 
fully explicit - from Eq. (\ref{equa4a}), $I_{n+1}$ in Eq. (\ref{equa4b}) can be written in terms 
of $I_n$ and $\theta_n$. In addition the map Jacobian has unitary norm, which guarantees 
the sympletic character of the theory. We also note that, in contrast to the standard case, the 
map displays strong nonlinear dependence on the wave amplitude. A linearized map, obtainable only 
when $\varepsilon \ll 1$, can be cast into the canonical form  
\begin{widetext}
\begin{eqnarray}
I_n= I_{n+1} - (1/\sqrt{2}) \varepsilon k \sqrt{I_{n+1}} \cos \theta_n \cos(k\,\sqrt{2 I_{n+1}} \sin \theta_n) \label{linear1}\\
\theta_{n+1} = \theta_n + {T \over \sqrt{1+2 I_{n+1}}}+{\varepsilon k \sin \theta_n \cos(k\,\sqrt{2 I_{n+1}} \sin \theta_n) \over \sqrt{8 I_{n+1}}}. 
\label{linear2}
\end{eqnarray}
\end{widetext}
The linear form can be naively obtained 
from the Hamiltonian (\ref{equa3aa}) alone if across the kick one assumes $\theta \rightarrow \theta_n$ and $I \rightarrow I_{n+1}$; 
our theory shows why these assumptions should be this way.


\section{Analytical and numerical results}

We are now in position to explore the properties of the fully nonlinear map (\ref{equa4a}) and (\ref{equa4b}).
When $\varepsilon \rightarrow 0$ the action $I$ becomes constant and the phase $\theta$ advances steadily according to 
\begin{equation}
\theta_{n+1} = \theta_n + {T  \over  \sqrt{1+2I}}.
\label{equa6}
\end{equation}
As mentioned earlier, we shall look into cases where both periods are similar, since a variety 
of cases fall in this ordering category. Let us further narrow our focus on the main resonance, the one for which 
each wave cycle corresponds to a full orbital gyration of the magnetized particles. This case corresponds to take 
$\theta_{n+1} - \theta_n = 2 \pi$, which sets the position of the main resonance at
\begin{equation}
I_{res} = {{T^2 - 4 \pi^2} \over 8 \pi^2}.
\label{equa7}
\end{equation}
Expression (\ref{equa7}) is only an estimate that does not include any effects resulting from the 
perturbing wave. Nevertheless, it provides a first step to investigate the role of resonances in the dynamical system. 
Recalling that the wave frequency is measured in units of the cyclotron frequency $q B_0/m$, when 
wave and cyclotron frequencies coincide, $T=2 \pi /1 = 2 \pi$ and $I_{res} \rightarrow 0$. For larger values of the 
wave period $T$, the resonance moves towards higher values of the action, and for smaller values the resonance cannot 
be realized for positive values of the action. Importantly, when relativity is suppressed with $I \rightarrow 0$ in 
Eq. (\ref{equa6}), the phase advance becomes independent of the action and one falls in a degenerate theory where 
the resonant island is absent \cite{Lichtenberg92}. This feature contrasts with the relativistic standard case where 
suppression of relativistic effects does not remove the main resonance \cite{Chernikov89, Nomura92}.

\begin{figure} [htb]
\begin{center}
\includegraphics[scale=0.7]{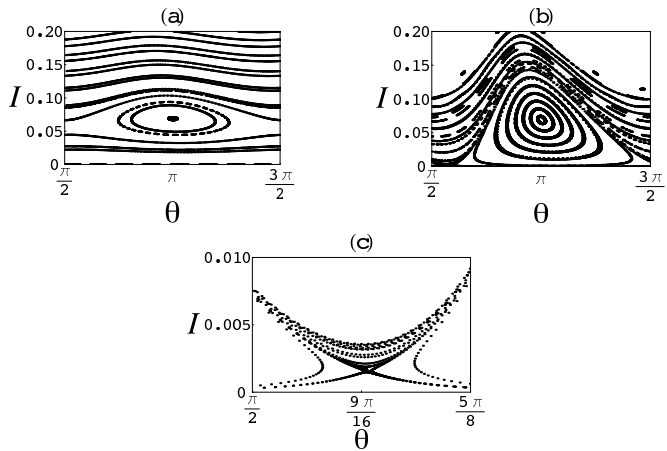}
\caption{Nonlinear resonance: pendulum-like with $\varepsilon k^2 = 0.1$ in (a), and $\varepsilon k^2 = 0.83$ 
in (b). Panel (c) details the positioning of the hyperbolic fixed point of panel (b) near $\pi/2$. 
$k=2$ and $T=2 \pi (1 + 1/15)$ in all cases.\label{fig1.eps}}
\end{center}
\end{figure}

We now look into a case where the resonance is present in the phase-space with the period $T$ slighly larger than the cyclotron 
period: $T = 2 \pi (1+1/15) > 2 \pi$. We also consider 
$k = 2$ as a representative wavevector of the modes analyzed, 
and display two distinct situations in Fig. \ref{fig1.eps}: $\varepsilon  k^2 = 0.1$ in panel (a) and $\varepsilon k^2 = 0.83$ 
in panel (b). It will become clear later why we work with the combination $\varepsilon k^2$ instead of the wave amplitude 
$\varepsilon$ alone.  In both cases the phase-spaces are fairly regular, but a remarkable fact distinguishes the panels apart. While in 
(a) the resonance takes the usual pendulum-like shape with elliptic and hyperbolic points 
approximately located at the same level along the action axis, in panel (b) one sees that the hyperbolic points move 
down to $I \approx 0$ while the elliptic point stays at the same original location. The relevance of this feature lies in 
the possibility of particle acceleration from low initial energies. Indeed, while in panel (a) particles launched with 
$I \approx 0$ remain with small values of the action, in panel (b) one sees that particles launched under the same 
condition $I \approx 0$ perform much larger excursions along the island separatrix.  Interestingly, the linearized map 
(\ref{linear1}), (\ref{linear2}) yields the angular position of the two hyperbolic points exactly at $\pi/2$ and $3 \pi /2$ 
for all values of $\varepsilon$.  Close examination of panel (b) shows however that this is only approximate, as one sees that 
both points sit to right of coordinates $\pi/2$ and $3 \pi/2$ respectively. The region near $\theta = \pi / 2$, in particular, 
is detailed in panel (c) where one can see the positioning of the hyperbolic point - numerical analysis reveals that 
it is located at $\theta = 1.13 \times \pi/2$, $I = 0.00157$. 
The intrincated nonlinear dependence of the full map (\ref{equa4a}), (\ref{equa4b})
on $\varepsilon$ displaces the hyperbolic points from their approximate phases, which is something to consider if one wishes 
to adjust wave-particle phase for optimum acceleration. 

Topology of Fig. \ref{fig1.eps} is similar in related contexts 
involving time dependent but spatially uniform perturbations \cite{Kim96}, although the fully 
analytical map (\ref{equa4a}), (\ref{equa4b}) reveals that the hyperbolic points in both cases do not coincide. 

Fig. \ref{fig1.eps}(b) also suggests that one way to improve the acceleration efficiency would be to 
increase the mismatch between cyclotron and wave frequency. According to the estimate of Eq. (\ref{equa7}), larger 
wave periods would pull the resonance upwards, and if one could choose a sufficiently large $\varepsilon$, the 
hyperbolic points could still be brought down to $I \approx 0$ enabling particles to coherently loop from small 
to large values of the action. There is however a natural limitation to this outlined procedure, which arises from 
the loss of stability of elliptic points as the wave amplitude $\varepsilon$ grows. In the case under study, one can 
expect the accelerating mechanism to work satisfactorily up to the point where the central point at $\pi$ looses 
stability. Beyond that, one can expect the intrusion of appreciable chaotic activity into the system with 
the concomitant loss of orbital coherence.  It is true that even before the elliptic point bifurcates chaos is present, but 
for the sake of simplicity we take the bifurcation as indicating the limits of regular regimes in the system. The condition 
we seek for regular acceleration can be thus stated as the one associated with periods $T$ for which one can 
maneuver the control parameters such that the hyperbolic points touch the axis $I=0$ {\it before} the elliptic
point bifurcates. Maximum, or optimum acceleration can be achieved for the particular period $T = T_{opt}$ where
touch down and bifurcation occur simultaneously.

To obtain the parameters for which hyperbolic ($h$) points reach $I=0$ we look for fixed points of Eq. (\ref{equa4b}) with 
periodicity $2 \pi$ at $I=0$, i.e., $\theta_{n+1} = \theta_n + 2 \pi$ with $\theta_h \equiv \theta_n ({\rm mod}\, 2 \pi):$ 
\begin{equation}
T+\arctan \left(\frac{\tan (\theta_h)}{1 + \frac{1}{2} \varepsilon k^2 \tan
   (\theta_h)}\right) = \theta_h + 2 \pi.
\label{equa4bfp}
\end{equation}
As mentioned earlier, fixed points of Eq. (\ref{equa4bfp}) are not aligned with $\theta = \pi/2, \> 3 \pi/2$ as in the 
linearized version of map (\ref{equa4a}) and (\ref{equa4b}). Nevertheless, careful inspection of Eq. (\ref{equa4bfp}) along with the use of trigonometric 
identities allows to find the following solution for the fixed points:
\begin{widetext}
\begin{equation}
\tan \theta_h = \frac{\varepsilon k^2 \tan (T) \pm \sqrt{\tan(T)}
   \sqrt{\varepsilon^2 k^4 \tan (T)+8 \, \varepsilon \, k^2-16 \tan
   (T)}}{2 \left(\varepsilon k^2-2 \tan (T)\right)}.
\label{equa8}
\end{equation}
\end{widetext}

Real solutions of expression (\ref{equa8}) exist only when $\varepsilon^2 k^4 \tan (T)+8 \varepsilon k^2-16 \tan(T) \geq 0$, 
from which one obtains the corresponding threshold for touch down ({\it td}): 
\begin{equation}
(\varepsilon k^2)_{td} =  {4 \over \tan(T)} \left(\sqrt{1 + \tan^2(T)} - 1\right).
\label{equa9}
\end{equation}

When $T=2 \pi (1 + 1/15)$, as used in Fig. \ref{fig1.eps} for instance, $(\varepsilon k^2)_{td}=0.85066$; this is why we 
chose that particular value of the product $\varepsilon k^2$ in Fig. \ref{fig1.eps} (b) - it is close to the touch down 
limit. 

As mentioned, one has regular acceleration only when touch down occurs before the central fixed point 
at $\theta = \pi$ looses stability via a period doubling bifurcation ({\it pd}). To obtain the conditions for period doubling 
we first of all observe that analysis of Eq. (\ref{equa4b}) at 
$\theta = \pi$ allows to conclude that result (\ref{equa7}) is in fact {\it exact}, independently of the value of $\varepsilon$. 
Then, linear stability analysis of Eq. (\ref{equa4b}) at the fixed point readily indicates that bifurcation occurs at:
\begin{equation}
(\varepsilon k^2)_{pd} = {4 \> T^2 \over \pi \> (T^2 - 4 \pi^2)}.
\label{equa10}
\end{equation}
We point out that from expressions (\ref{equa9}) and (\ref{equa10}) it becomes apparent that the effective 
parameter controlling the perturbative strength is the combination $\varepsilon k^2$ and not the wave amplitude alone, 
as commented earlier.

\begin{figure} [htb]
\begin{center}
\includegraphics[scale=0.7]{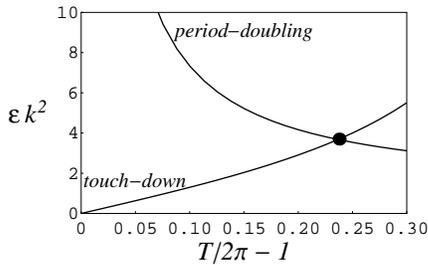}\\
\caption{Comparison of the threshold curves for period-doubling and touch down. Labels 
respectively indicated. The black dot indicates the point of optimum acceleration.\label{fig2.eps}}
\end{center}
\end{figure}

The next step is to compare both thresholds furnished by Eqs. (\ref{equa9}) and (\ref{equa10}). This is 
done in Fig. \ref{fig2.eps} where one sees that there is indeed a critical point in the plane $\varepsilon k^2$ versus $T$, 
where touch down and period doubling occur simultaneously. For values of the period $T$ below the one at the critical 
point, touch down occurs earlier than doubling, and after the critical point the ordering is reversed. 
At the critical point one attains optimum conditions for regular acceleration, as discussed earlier. The critical 
point period can be numerically evaluated as $T_{opt} \approx 1.24 \times 2 \pi$. We also have 
$(\varepsilon k^2)_{opt} = 3.67843$ and $I_{opt}=0.26469$. This means that the dimensionless 
momentum excursion reads $p_x \sim \sqrt{2 (2 I_{opt})} \sim 1.03$ which provides a significant amount 
of acceleration $v_x = p_x/\sqrt{1+p_x^2} \sim 0.72$ especially when one looks at ion dynamics. To obtain the estimate 
for maximum momentum we simply double the value of the action at the fixed point, as suggested by the topology of 
Fig. \ref{fig1.eps} (b), and make use of expression (\ref{equa7}) along with the relation between action-angle 
and cartesian coordinates. $\varepsilon_{opt} \sim 1$, so we are beyond the validity of linear approximation 
(\ref{linear1}), (\ref{linear2}).

\section{Conclusions}

To conclude, we developed and investigated an analytical canonical map describing the relevant case of 
impulsive interaction of waves and magnetized 
relativistic particles. In contrast to the better known standard canonical case, the present map is shown to be nonlinear 
in the wave amplitude, which affects the positioning of fixed points. Also, special regular orbits were shown to exist 
which can be used to accelerate particles. These special orbits bifurcate to chaos when the interaction parameter 
grows beyond a certain threshold. The orbits and all the associated bifurcation process are only present when 
nonlinear relativistic mass correction of particles is properly taken into account. If relativistic effects 
are erroneously ignored, resonant islands are replaced with KAM curves \cite{Kim96}. 

Comparing ours with similar settings, we initially recall that topology of Fig. \ref{fig1.eps} resembles that of related contexts 
involving time dependent but spatially uniform perturbations \cite{Kim96}, although the fully 
analytical map (\ref{equa4a}), (\ref{equa4b}) reveals different positioning of fixed points.
We point out as well that relativistic generalizations of the Karney problem applied to the interaction of magnetized electrons and 
perpendicular electrostatic lower hybrid waves have also been studied, with closer focus 
either on pendulum-like islands or on waves with much larger than the relativistic cyclotron frequencies \cite{Rax91}. 
Our problem here is more directed to the resonant interaction and generation of the large distorted island. 

\begin{acknowledgments}
We acknowledge support by CNPq (Brazil) and AFOSR (USA) under grant number FA9550-09-1-0283.
\end{acknowledgments}

%
%
%
%

\end{document}